\documentclass[11pt,english,twoside]{article}
\pdfoutput=1
\usepackage{jheppub}
\usepackage{lscape}
\usepackage{amsmath}
\usepackage{slashed}
\usepackage{mathtools}
\usepackage{smartdiagram}
\usepackage{metalogo}
\usepackage{tikz,tikz-3dplot}
\usepackage{tikz-cd}
\usepackage[framemethod=TikZ]{mdframed}
\usepackage{pdfpages}

\definecolor{cof}{RGB}{219,144,71}
\definecolor{pur}{RGB}{186,146,162}
\definecolor{greeo}{RGB}{91,173,69}
\definecolor{greet}{RGB}{52,111,72}

\def\eps{\epsilon}

\def\-{\hphantom{-}}

\def\s2{\frac{1}{\sqrt2}}

\def\beqa{\begin{eqnarray}}
\def\eeqa{\end{eqnarray}}

\def\eps{\epsilon}

\def\eps{{\epsilon }}

\def\mg{m_{3/2}}
\def\mg2{m^2_{3/2}}

\def\Dsl{\,\raise.15ex\hbox{/}\mkern-13.5mu D} 

\newcommand{\eq}[1]{\begin{equation}
                     \begin{split} #1 \end{split}
                     \end{equation}}

\begin{document}

\title{Inflationary implications of the Covariant Entropy Bound and the Swampland de Sitter Conjectures}

\author[a]{Dibya Chakraborty,}
\author[b]{Cesar Damian,} 
\author[a]{Alberto Gonz\'alez Bernal}
\author[a]{and Oscar Loaiza-Brito}

\affiliation[a]{Departamento de F\'isica, Universidad de Guanajuato, \\
Loma del Bosque No. 103 Col. Lomas del Campestre C.P 37150 Leon, Guanajuato, Mexico.}
\affiliation[b]{Departamento de Ingenier\'ia Mec\'anica, Universidad de Guanajuato, \\
Carretera Salamanca-Valle de Santiago Km 3.5+1.8 Comunidad de Palo Blanco, Salamanca, Mexico}

\emailAdd{dibyac@fisica.ugto.mx}
\emailAdd{cesaredas@fisica.ugto.mx}
\emailAdd{ja.gonzalezbernal@ugto.mx}
\emailAdd{oloaiza@fisica.ugto.mx}

\date{\today}

\abstract{We present a proposal to relate the de Sitter Conjecture (dSC) to the Covariant Entropy Bound (CEB). By assuming an early phase of accelerated expansion where the CEB is satisfied, we take into account a contribution from extra-dimensions to the four-dimensional entropy which restricts the values of the usual slow-roll parameters. We show in this  context that the dSC inequalities follow from the CEB  $-$including their mutual exclusion$-$ in both single and multi-field inflationary scenarios. We also observe that  the order one constants, $c$ and $c'$  in the conjecture are given in terms of physical quantities such as the change in entropy over time,  the Hubble constant and the dynamics of the effective scalar fields. Finally, we give a simple example to illustrate a possible contribution to the four-dimensional entropy from a flux string scenario.\\
}
\arxivnumber{}


\maketitle


\section{Introduction}
The Swampland program has reached great advances in the last few years in its pursuit to characterize those features distinguishing effective theories that can be consistently completed into a quantum gravity theory (Landscape) from those which do not (Swampland)\cite{Vafa:2005ui, Ooguri:2006in, Palti:2019pca, vanBeest:2021lhn}. The boundary between the Landscape and Swampland is usually defined by a series of bounds on quantities of the proper effective theory which in turn are expressed in terms of Planck mass $M_p$. In the limit where gravity decouples, i.e., $M_p\rightarrow \infty$,  the Swampland constraints vanish.  A question to be answered is whether these boundaries can be traced back to some essential microscopic physics in the Quantum Gravity Theory, or accordingly,  to some fundamental constraints in the 10-dimensional String Theory formulation.\\

Many different bounds have been proposed as Swampland Conjectures, although it is expected that all of them are related in some way
through more fundamental principles (see review \cite{Palti:2019pca} and lectures \cite{vanBeest:2021lhn} for references). Some of them  are based on solid grounds  as the Distance Conjecture and the Weak Gravity Conjecture, while others, as the de Sitter Conjectures,  were initially motivated by some  {\it empirical} evidence based on string models \cite{Danielsson:2018ztv, Obied:2018sgi}, and appropriately refined into its final form in order to be  compatible with some well known effective scenarios such as the Higgs potential \cite{Denef:2018etk} among others.  The refined version of the conjecture was motivated by stability of de Sitter extrema \cite{Andriot:2018wzk}, by cosmological arguments through the slow-roll parameter $\eta$ \cite{Garg:2018reu} as well as by entropy arguments on the dS space \cite{Ooguri:2018wrx}. Many logic and solid arguments have been followed in the literature \cite{Andriot:2018ept, Conlon:2018eyr, Dasgupta:2018rtp, Andriot:2018mav, Geng:2019zsx, Atli:2020dni, Dasgupta:2019vjn, Blumenhagen:2019kqm, Damian:2019bkb, Andriot:2020vlg} so far with the purpose to give stronger ground basis to this conjecture.  A matter of particular interest to us is the series of arguments supporting the de Sitter Conjecture (dSC)  based on the laws of thermodynamics \cite{Seo:2019mfk, Gong:2020mbn, Blumenhagen:2020doa, Kobakhidze:2019ppv}. Relating the dSC to thermodynamic principles, such as a positive temperature phase and the concavity of  the entropy functional,  respectively have also been  considered in the literature. Furthermore, as stochastic effects become relevant, it is found by thermodynamic arguments that  slow-roll conditions either violate the second law or the swampland dSC \cite{Wang:2019eym,Brahma:2019iyy}.\\

In its original form, the de Sitter Conjecture states that
\eq{
\frac{|\nabla V|}{V} \geq c \,\qquad\text{or} \qquad \frac{\text{min} \nabla_i \partial_j V}{V} \leq -c' \,,
\label{SdSC}
}
with $c$ and $c'$  positive constants of order one in Planck units and $V$ the effective scalar field potential.
Besides the well known implication of discarding the existence of  dS minima, dSC also excludes the presence of inflation driven by a single field \cite{Agrawal:2018own}. However, as shown in \cite{Achucarro:2018vey}, dSC seems to be compatible with a multi-field inflationary scenario. \\

Recently, cosmological implications of the dSC have been proven to be a matter of great interest  \cite{Heisenberg:2018yae, Heisenberg:2018rdu, Akrami:2018ylq, Wang:2018duq, Chiang:2018lqx, Matsui:2018bsy,  Schimmrigk:2018gch, Agrawal:2018rcg, Damian:2018tlf, Trivedi:2020xlh, Trivedi:2020wxf, Das:2020xmh, Brandenberger:2020oav}. Roughly speaking we could say that there are actually two scenarios: the first one concerns the study of the so called late-acceleration of the Universe, to which the dSC seems to choose a Dynamical Dark Energy mechanism over a cosmological constant related to a dS minimum.  A prototypical example of such a model is Quintessence \cite{Cicoli:2018kdo, Blumenhagen:2018hsh,Marsh:2018kub, Olguin-Tejo:2018pfq, Ibe:2018ffn, Brahma:2019kch, Cicoli:2020noz}; the second one focuses on the study of the early inflationary stage, prior to reheating, where the simplest model involves a field slowly rolling down an almost flat scalar potential. As mentioned above,  the dSC seems to discard inflation driven by a single scalar field. \\

It is then interesting for our purposes to review an (heuristic) argument supporting the dSC in both scenarios above mentioned. For an accelerated expansion, the  dSC is  expected by taking into account how entropy increases over distance in a de Sitter space \cite{Palti:2019pca}. Essentially it invokes the association between an entropic system with the temperature given to  the  dS Apparent Horizon. The number of microstates of the corresponding Hilbert space contributes to the entropy which increases with the field distance. Since it is possible to estimate that the entropy goes like the inverse of the cosmological constant,  one can conclude that the latter cannot be a constant after all. Since  it is conjectured that the mass of the field exponentially decreases with field distance, one concludes that $V'\sim cV$. \\

In contrast to late acceleration, during the early inflationary stage of the Universe, there is almost no  contribution to entropy, a reason being that during an accelerated expansion,  the comoving entropy density remains constant \cite{Bousso:2002ju}.  This is true if one considers the 4-dimensional Universe a closed system, but opens up the possibility to an inflow of entropy from extra dimensions in a string theory scenario. \\

In this paper we are interested in studying the implications of the CEB \cite{Bousso:2002ju, bousso1999covariant, Bousso:2015eda, Bousso:2014sda} $-$under the assumption of an entropy inflow  from extra dimensions$-$  into a four-dimensional expanding Universe.\\

 A direct outcome of relating the CEB to the dSC is the construction of explicit expressions for the involved constants $c$ and $c'$ in the dSC innequaities (\ref{SdSC}) in terms of physical quantities,  among which time variations of entropy play a leading role. We also obtain that the exclusion between the two inequalities in the dSC is a natural and direct  consequence of the CEB under the assumptions stated above. Therefore, the validity of the dSC relies on the actual value of the entropy change for which it would be necessary to compute the entropy contribution from a specific string compactification scenario. A simple example concerning a toroidal flux compactifcation is presented in this context.\\

However, it is important to emphazise that we are not generating an expanding (inflationary or not) four-dimensional Universe through fluxes. The source of an expanding Universe from a string theory point of view is beyond the scope of this paper.  Our interest focuses on the compatibility of an inflationary Universe with the CEB\footnote{The source of entropy in extra dimensions may rely on some basic and fundamental quantities in string theory  such as the presence of NS-NS and R-R fluxes. They play a crucial role in many phenomenological applications such as generating the correct hierarchies of scales in four-dimensional effective field theories by warping the geometry and particularly in inflationary model building \cite{Grana:2005jc, Blumenhagen:2017cxt, Blumenhagen:2015xpa, Blumenhagen:2015kja, Blumenhagen:2014nba} among others. The Tapole constraints also lead to important implications on effective scenarios  \cite{CaboBizet:2019sku, Betzler:2019kon, Sati:2019tqq, CaboBizet:2020cse, Plauschinn:2020ram, Bena:2020xrh}.}.\\

The paper is organized as follows. In section 2 we briefly review some basics regarding the CEB and establish a relation between the change of entropy over time and the slow-roll parameter $\epsilon_H$ (Eq. \ref{gamma}). In section 3 we use this relation to study its implications on single field and multi-field inflationary scenarios. Our main result is that we are able to reproduce the inequalities of the dSC as well as the mutual exclusion between them in terms of the time variation of entropy sourced from physics in extra dimensions as well as the explicit form of the constants $c$ and $c'$.  A simple example concerning a toroidal flux compactification is presented to show how fluxes contribute to the four-dimensional entropy.  Finally we discuss our conclusions and some final comments.

\section{Covariant Entropy Bound and Inflation}

The Covariant Entropy Bound (CEB) is a well tested conjecture in many gravitational scenarios including cosmology \cite{bousso1999covariant, Bousso:2002ju, Bousso:2014sda, Bousso:2015eda}.  Basically, it states that for  a codimension 2 surface $B$, there exist at least two light-sheets $L$, defined as the codimension 1  null hypersurfaces orthogonal to $B$ spanned by light-type trajectories terminating at a singularity or a caustic,  with a negative expansion rate $\theta$. These light-sheets can be future or past-directed. The general statement is that the entropy associated to the light-sheet  $S(L[B])$, at a given time, is bounded by the area $A$ of the surface $B$ as\footnote{ In terms of universal constants, $S\leq \frac{c^3\kappa_B}{4\hbar G} A$. In this paper we take all constants -including $M_{\text{Pl}}$ to be equal to 1.}
\begin{equation}
S(L[B])\leq \frac{A(B)}{4}.
\end{equation}
In a flat 4-dimensional expanding Universe described by the  Friedmann-Robertson-Walker (FRW) metric 
\eq{
ds^2_4=-dt^2+a^2(t)(dr^2+r^2d \Omega^2),
\label{FRW}
}
for a given surface $B$ with area $A$ at a time $t$, there are 4 different null-hypersurfaces $L$, which can be past(future)-directed and are labeled by their in(out)-going direction with expansion rates denoted by $\theta_{(\pm,\pm)}=\pm\dot{a}\pm(1/r)$. In this scenario, it is possible to define the Apparent Horizon (AH) as the surface $B$  with at least two null expanding rates, specifically with $\theta_{+-}=\theta_{-+}=0$. As shown in \cite{Bousso:2002ju}, this is given by defining the radius of the AH at a time $t$ as $r_{AH}=(aH)^{-1}$, with $H$ the Hubble constant defined as $H=\dot{a}/a$, where $\dot{a}$ denotes the time derivative with respect to the proper time $t$. As the Universe expands,  the comoving Hubble radius $r_{AH}$ shrinks and the AH  comoving  area given by
\begin{equation}
A_{AH}=4\pi a^2 r_{AH}^2=\frac{4\pi^2}{H^2}
\end{equation}
increases, as expected by a negative  $\dot{H}$. Notice that the corresponding past directed light-sheet has an expansion rate $\theta_{--}=-2\dot{a}$ which is negative for $\dot{a}>0$. \\

For any area, bigger or smaller than $A_{AH}$, it is conjectured that the CEB will be always fulfilled. In particular, since we are considering an expanding Universe, smaller areas are past-directed and the corresponding light-sheets are represented by the expansion rate $\theta_{-,\pm}$. Hence, for $t_1<t_2$, the area $A(t_2)$ is connected to a smaller area $A(t_1)$ by a light-sheet with expansion rate $\theta_{-,\pm}$ which is referred to as an anti-trapped surface \cite{Bousso:2002ju}. The CEB can be strengthened by considering the entropy associated to the light-sheet between the surfaces $A(t_2)$ and $A(t_1)$, as shown in \cite{Bousso:2014sda}, with the entropy change bounded as $\Delta S\leq (A(t_2)-A(t_1))/4$, and where $\Delta S$ is thought to be the von Neumann entropy related to the difference between matter associated to the light-sheets and the entropy of vacuum. \\

We are interested in applying this bound for the AH area in an expanding flat FRW Universe.  One can wonder about the entropy production between  some time interval during inflation. 
By assuming an adiabatic expansion, the comoving entropy density $s=dS/dV$ is constant, indicating that there is no entropy variation. This can be understood as a consequence of considering the Universe as a closed system in which the production of matter has not yet occurred (reheating) and therefore there is no  variation of entropy over time in contrast to the AH area. However, we can think that in the framework of a quantum theory of gravity there should be extra contributions to the entropy allowing a condition of the form $\dot{S}
\equiv \frac{dS}{dt}\ne 0$. This can actually  happen in the presence of extra dimensions as in string theory. \\
	
Following  \cite{Bousso:2014sda}, let us consider the AH surface at different times, $t_2$ and $t_1$ with $t_2> t_1$. We then have  the entropy variation satisfies the bound
\eq{
\Delta S\leq \frac{A_{AH}(t_2)-A_{AH}(t_1)}{4},
\label{derivativeCEB}
}
where $\Delta S$ is the difference between the entropy at time $t_2$ and $t_1$.  Therefore,  an infinitesimal variation  of time helps us to rewrite Eq.(\ref{derivativeCEB}) in the form\footnote{ Although the CEB was proved to  be consistent only for scenarios in which the Null Energy Condition (NEC) holds,  it was shown in  \cite{Bousso:2014sda} that Buosso bound can also be satisfied independently of the NEC. }

\begin{equation}
\gamma\equiv\frac{H \dot{S}}{2\pi}\leq \epsilon_H,
\label{gamma}
\end{equation}
with $\epsilon_H=-\dot{H}/H^2$ is the usual slow-roll inflationary parameter.\\

Notice that this holds even for non-constant $H$. For the specific case in which we have an exponential expansion, $\epsilon_H$ must be much smaller than 1, and in consequence $\gamma$ must also be very small.  Slow-roll inflation is then ruled out if $\gamma\sim {\cal O}(1)$ or for $\gamma\gg 1$.\\

In the following we shall assume that a variation of entropy in time comes from extra dimensions within a string theory based scenario. In this context we expect that entropy must be a function of the Hubble parameter $H$, time and other string parameters which we shall ignore to our present purpose. \\

It is the goal of this work to study the implications of a non-zero gamma (Eq.(\ref{gamma})) on the formulation of  the dSC.

\section{Inflation and Effective Field Theory}
In this section, we study the implications of assuming a non-zero flux entropy $\gamma$ in an effective field theory involving scalar fields incorporating both  single and  multi-field scenarios.\\

\subsection{Single field inflation}
From the inequality (\ref{gamma}) and the usual Friedmann-Robertson-Walker (FRW) equations of motion, we obtain 
\eq{
\gamma\leq\epsilon_H=
\frac{1}{2}\left(\frac{A+\frac{\partial_\phi V}{V}}{1+B}\right)^2\\
\label{eH}
}
where  $A=\frac{\ddot{\phi}}{V}$ and $B=\frac{\dot{\phi}^2}{2V}$. Then one can immediately write
\eq{
\left|\frac{\partial_\phi V}{V}\right|\geq |1+B|\sqrt{2\gamma}-|A|\equiv c.
}
The value of $c$ clearly depends on different scenarios, which we now proceed to describe:
\begin{enumerate}
\item 
$\gamma\ll 1$: this means that  $\gamma\leq \epsilon_H\ll 1$. This implies that the factor $(1+B)$ must be much greater than $A$ for $\partial_\phi V/V\ll 1$. This allows a very small value for $c$ and the usual identification $\epsilon_H\approx\epsilon_V$.  It follows that an accelerated expansion is a possibility. This case also involves the presence of  slow-roll inflation  by taking $A\sim B\sim 0$.  However, it also allows values of $c$ to be of order 1 or bigger independently of $\gamma$ (for adequate values of $A$ and $B$).
\item 
$\gamma\geq 1$:  if $\gamma$ is of order one or bigger, $\epsilon_H$ cannot be sufficientely small to allow an inflationary stage of the Universe. The value of $c$ can also be smaller, bigger or of order 1. Under the usual slow-roll conditions, $c=\sqrt{2\gamma}$ and its value is determined absolutely by the entropy variation during expansion. If $\gamma$ is of order 1 or bigger, inflation is not allowed and the first inequality in the dSC is obtained as
\eq{
|\partial_\phi V|\geq \sqrt{2\gamma}~|V|.
\label{dSCsingle}
}
Again the outcome is that an accelerated expansion and inflation are not allowed.
\end{enumerate}
We shall come back to these points once we consider the eigenvalues of the Hessian matrix which will allow us  to observe if we can reproduce the mutually exclusive dSC  (\ref{SdSC}).\\

 Let us now be emphatic on two important aspects: a compatible inflationary scenario requires at least $\gamma<\epsilon_H < 1$; in a closed 4-dimensional Universe $\gamma$ is almost zero since  entropy is almost constant during expansion \cite{Bousso:2002ju}, and we have 
\eq{
\epsilon_H\approx\epsilon_V\equiv\frac{1}{2}\frac{|\partial V|^2}{V^2}\geq 0,
\label{eHeVsingle}
}
allowing  $\epsilon_V<1$ as well. Inflation can be translated into conditions on the effective scalar potential for an almost null $\gamma$. \\


\subsection{Multi-field inflation}
Let us generalize the above results to an $N$-field scenario. As shown in \cite{Pinol:2020kvw}, it is more convenient to switch to the kinematic $-$Frenet-Serret$-$ frame to compute the equations of motion of an effective theory in the presence of a scalar potential $V=V(\phi_a)$ with $a=1,\dots N$.  See Appendix \ref{A1} for more details.\\

The equations of motion can be written in terms of the unitary tangential vector $\hat{e}_1$, the normal $\hat{e}_2$ and the $(N-2)$ binormal $\hat{e}_i$, as
\begin{equation}
(\ddot\phi+3H\dot\phi+V_1)\hat{e}_1+(-\dot\phi\Omega_1+V_2)\hat{e}_2 +V_i\hat{e}_i=0,
\label{eomFS}
\end{equation}
with $i=3,\dots, N$ and $\dot\phi$ is the norm of the vector field $\vec{\dot\phi}=(\dot\phi^1,\dots ,\dot\phi^n)$. Since the Frenet-Serret frame $\{\hat{e}_1, \hat{e}_2, \hat{e}_i\}$ is an orthonormal basis, we find that
\begin{eqnarray}
\ddot\phi+3H\dot\phi+V_1&=&0,\\
-\dot\phi\Omega_1+V_2&=&0,\\
V_i&=&0.
\end{eqnarray}
Notice all projections of $\vec{\nabla}V$ along normal vectors $\hat{e}_i$, denoted by $V_i$, vanish for $i=3,\dots,n$, implying that we can focus only on the field trajectory on the osculating plane. \\

Now, together with the Einstein-Friedmann equation 
\eq{
3H^2=\frac{\dot{\phi}^2}{2}+V,
}
and by the usual definitions of the slow-roll parameters, 
\eq{
\epsilon_V\equiv\frac{1}{2}\left(\frac{\vec{\nabla}V}{V}\right)^2=\tilde{\epsilon}_V+\hat{\epsilon}_V,
}
with 
\begin{equation}
\tilde{\epsilon}_V=\frac{1}{2}\left(\frac{V_1}{V}\right)^2,\quad \hat{\epsilon}_V=\frac{1}{2}\left(\frac{V_2}{V}\right)^2,
\end{equation}
we can construct a relation between $\epsilon_H$ and $\hat\epsilon_V$  by

\begin{equation}
\epsilon_H^2-(6+\alpha)\epsilon_H+9=0,
\label{ealpha}
\end{equation}
where 
\eq{
\alpha=\frac{\Omega_1^2}{H^2\hat\epsilon_V}=\frac{2V^2}{H^2\dot{\phi}^2},
}
with $\hat{\epsilon}_V \ne 0$. Notice that $\alpha\rightarrow 0$ implies $\epsilon_H\rightarrow 3$ for which an accelerated expansion $-$and therefore inflation$-$ is ruled out. 
An almost zero $\alpha$ means two things: one is that the kinetic energy is much bigger than $V$;
 the second possibility involves $\hat{\epsilon}_V$ growing faster than the ratio $\Omega_1/H$. \\

\noindent
Let us revisit the inequality (\ref{gamma}).
By using $\epsilon_H$ as a function of $\hat\epsilon_V$ from (\ref{ealpha}) we obtain:
\begin{eqnarray}
\gamma&\leq&\left(3+\frac{\alpha}{2}\right)\pm\sqrt{\left(3+\frac{\alpha}{2}\right)^2-9},
\label{gammaev}
\end{eqnarray}
which provides us different implications for the following two possible scenarios:
\begin{enumerate}
\item
$\gamma\geq 3+\frac{\alpha}{2}$: since $\alpha\geq 0$, $\epsilon_H>1$ and  an accelerated expansion is  discarded.  It follows from Eq.(\ref{gammaev}) that
\begin{equation}
\hat{\epsilon}_V\leq \frac{\Omega_1^2}{H^2}\frac{\gamma}{(\gamma-3)^2}.
\end{equation}
We observe that $\gamma$ can take values from 3 to $\infty$.  However,  in this case it is not possible to fix a lower bound for $\epsilon_V$.
\item
$\gamma\leq 3+\frac{\alpha}{2}$: from Eq.(\ref{gammaev}) we have 
\begin{equation}
\hat{\epsilon}_V\geq 
\frac{\Omega_1^2}{H^2}\frac{\gamma}{(\gamma-3)^2}.
\label{gammainn}
\end{equation}
Then a lower bound for $\epsilon_V$ follows in the form of
\begin{eqnarray}
\frac{1}{2}\frac{|\nabla V|^2}{|V|^2} &=& \widetilde\epsilon_V+\hat\epsilon_V,\nonumber\\
&\geq& \widetilde\epsilon_V+\frac{\Omega_1^2}{H^2}\frac{\gamma}{(\gamma-3)^2}.
\end{eqnarray}
From $\epsilon_H=-\dot{H}/H^2$ and from the fact that $\dot{V}=V_1\dot\phi$, we obtain
\begin{equation}
\widetilde\epsilon_V=\frac{1}{2}\left[ A+\sqrt{2\epsilon_H}(1+B)\right]^2,
\end{equation}
with $A$ and $B$ defined in Eq.(\ref{eH}). This result together with 
\begin{equation}
\epsilon_H=\frac{\dot{\phi}^2}{2H^2}=\frac{V^2_2}{2H^2\Omega_1^2},
\end{equation}
implies that
\begin{eqnarray}
\frac{|\nabla V|^2}{|V|^2} &\geq&\left( A+ \frac{V_2}{H\Omega_1}(1+B)\right)^2+\frac{\Omega_1^2}{H^2}\frac{2\gamma}{(\gamma-3)^2},
\label{first}
\end{eqnarray}
from which the parameter $c$ of the first inequality of (\ref{SdSC}) reads
\begin{equation}
c^2=\left( A+ \frac{V_2}{H\Omega_1}(1+B)\right)^2+\frac{\Omega_1^2}{H^2}\frac{2\gamma}{(\gamma-3)^2}.
\label{cmulti}
\end{equation}
 In this case, $\gamma$ can be smaller than 1 and inflation is in principle not discarded, indicating the fact that the dSC is fulfilled if $c $ is of order 1. 
\item
Notice that, in the slow-roll limit, we have  $ \epsilon_H\sim\widetilde\epsilon_V$, which implies that $A\sim B\sim 0$, 
 inflation seems to be allowed for $\gamma\sim \epsilon_H$ since
\begin{equation}
\frac{|\nabla V|^2}{|V|^2} \geq 2\gamma\left(1+\frac{\Omega_1^2}{H^2}\frac{1}{(\gamma-3)^2}\right).
\label{AB=0}
\end{equation}
For $\gamma\ll 1$  it is straightforward to see that the above expression reduces to
\begin{equation}
\epsilon_V\sim \epsilon_H\left(1+\frac{\Omega_1^2}{9H^2}\right).
\end{equation}
Therefore, in a multi-field scenario, inflation is allowed with $\epsilon_V$ of order 1, as long as $\gamma$ is very small.  This is precisely the statement by Ach\'ucarro and Palma in \cite{Achucarro:2018vey}. A stringy example in which it is possible to fulfill the Swampland Constraint in a multi-field scenario by considering fat inflatons was studied in \cite{Chakraborty:2019dfh}.
\end{enumerate}

Observe that  expression (\ref{cmulti}) allows us to determine the physical origin of the constant $c$ and its direct relation, under our initial anzatz,  to a quantum theory of gravity from a 4d effective field theory point of view. \\

\subsection{Refined Swampland de Sitter Conjecture}

Now, let us try to reproduce the second inequality in the dSC (\ref{SdSC}) and determine whether both of those inequalities are mutually exclusive or not.  First of all, let us assume that $\gamma$ is defined in a way such that we can write
\eq{
\dot\gamma\leq \mathcal{O}(1)\dot\epsilon_H.
}
This is a reasonable assumption since $\gamma$ must be smaller than $\epsilon_H$ during a finite time interval in which an exponential accelerated expansion occurs. Since we are assuming $\gamma/\epsilon_H\leq 1$, we can restrict to a time interval during which  $\dot\gamma/\gamma\leq \dot\epsilon_H/\epsilon_H$ and then
 \eq{
 D_t \log \gamma \leq D_t \log \epsilon_H \,.
\label{Dlog}
 }
The left hand side of Eq.(\ref{Dlog}) can be written in terms of  $\eta_H=-\ddot{H}/(2H\dot{H})$ and $\epsilon_H$, as
\eq{
 -\frac{1}{2H} D_t \log \gamma\geq \eta_H - \epsilon_H.
\label{in1}
 }
On the other hand, by taking the derivative of the Einstein-Friedmann equations of motion,
 \eq{
 \eta_H + \epsilon_H =  \frac{D_t V_1}{3 H^2 \dot \phi}  + \omega' ,
 }
for $\omega' = \dddot \phi / 3\dot \phi H^2$ (this expression is consistent with the well known result, $\eta_H+\epsilon_H \approx \eta_V$ for  $\eta_H = - \frac{\ddot \phi}{H \dot \phi}$ and $\eta_V = \frac{D_t V_1}{3 H^2 \dot \phi}$). Thus,  inequality (\ref{in1}) can be written as
\eq{
\frac{D_t V_1}{3 H^2 \dot \phi}  \leq  -\frac{1}{2H} D_t \log \gamma + 2 \epsilon_H - \omega' .
}
Now, since
\eq{
\frac{D_tV_1}{3H^2\dot\phi}=\frac{Q_1-\Omega_1^2}{3H^2}
}
where $Q_1$ is a quadratic form given by
\eq{
Q_1=V_{ab}e^a_1e^b_1,
}
with $e^a_1=\dot\phi^a/\dot{\phi}$ (see Appendix \ref{A1}) it follows that
\eq{
Q_1\leq -\frac{3H}{2}D_t\log{\gamma}+\Omega_1^2 +3H^2(2\epsilon_H-\omega').
}
Under the condition of orthonormality $\sum_a|e^a_1|^2=1$, it is straightforward to show that the quadratic form $Q_1$ has a minimum (maximum) value given by the smallest (largest) eigenvalue $\lambda_{min}$ ($\lambda_{max}$) of the Hessian matrix $V_{ab}$ in the basis $\{e^a_i\}$. Hence we obtain

\eq{
\lambda_{min}\leq  -\frac{3H}{2}D_t\log{\gamma}+\Omega_1^2 +3H^2(2\epsilon_H-\omega').\\
\label{lambda}
}

Let us now stress the point that inflation claims that $\gamma\leq\epsilon_H\ll 1$ with $A\sim B\sim 0$ and in turn $\epsilon_H\sim\widetilde\epsilon_V$. Under these circumstances we have the following options:
\begin{enumerate}
\item
If $\Omega_1^2<H^2$ then from (\ref{AB=0})
\eq{
\frac{|\nabla V|^2}{|V|^2}\geq 2\gamma=c,
\label{1des}
}
where $c$ is not of order 1.  However, since $\gamma\ll 1$ we also have $\frac{1}{H}D_t\log\gamma \gg 1$ for $H$ not very small. Furthermore, using the inequality  $HD_t\log\gamma>\Omega_1^2$ and 
Eq. (\ref{lambda}) we can write
\eq{
\frac{\lambda_{min}}{V}\leq -\frac{1}{2H} D_t\log\gamma\equiv -c'.
\label{2des}
}
Observe that, while the first inequality of the dSC is not fulfilled, there is no obstruction for the second one to be satisfied. This also includes the case with $\Omega_1=0$, corresponding to a single scalar field scenario.
\item
If $\Omega_1^2>H^2$ it follows from (\ref{AB=0}) that $c$ can be of order 1. On the other hand, in this limit
there is always a value for $H$ such that  $D_t\log\gamma\gg H$. Then it is possible that $\Omega_1^2$ and $HD_t\log\gamma$ are of the same order, implying that the difference is not necessarily negative. Therefore, while the first inequality in (\ref{SdSC}) can be fulfilled by an appropriate value of $\Omega_1$, the second one is compromised. 
\end{enumerate}
Notice that, under the slow-roll conditions and with $\gamma\ll 1$, our two inequalities (\ref{first}) and (\ref{lambda}) are mutually exclusive as proposed in the original dSC while inflation is still  allowed. This depends clearly on the value of $\gamma$ which according to our anzatz, is different from zero due to quantum gravity effects, also in accordance with the Swampland conjectures.\\

Notice that for $\gamma \sim  c V^{n-2/4}$ for some $n$, the bound (\ref{1des}) limits the presence of eternal inflation \cite{Blumenhagen:2020doa}.\\

\subsection{Lyth's bound and the Swampland Distance Conjecture}
It has been found that, for a canonical single-field slow-roll model of inflation, generally  the  overall  field  displacement  $\Delta\varphi$ experienced  by  the  inflaton  during  the  quasi-de  Sitter phase must satisfy a lower bound, which is known as the \textit{Lyth bound} 
\eq{
\Delta\phi\simeq \sqrt{2\eps_H}\Delta N,
\label{use1}
}
where $\Delta N$ is the effective number of efolds during inflation needed to be at least  60. So, we can rewrite (\ref{use1}) as,

\eq{
\Delta\phi\simeq \sqrt{2}\cdot 60\sqrt{\eps_H}=\kappa\sqrt{\eps_H},
}
where $\kappa$ is a number of $\mathcal{O}(10)$. Hence, during inflation field displacement is expected to be of order 1 in Planck units, although larger orders have been considered in the literature and known as trans-Planckian distances. However, recently it has been conjectured (see \cite{Palti:2019pca, vanBeest:2021lhn} and references therein) that field displacements have an upper bound, probably connecting to other Swampland Conjectures, as the Distance Conjecture.\\

Now, in the case of multifield inflation, the field excursion $\Delta \phi$ has an extra contribution coming from non-trivial angular motion $\Omega_1$ and entropy mass  $M$ \cite{Achucarro:2018vey}, 
\eq{
\Delta\phi\gtrsim  \kappa\sqrt{\frac{\eps_H}{\beta}}
\label{use2}
}
where $\beta$ is a function depending on $\Omega_1$ and $M$. Now following \cite{Achucarro:2018vey}, if the fluctuating mode crosses the horizon in the linear regime, i.e., if the condition $(1-c_s^2)H<c_s M$ is satisfied, then $\beta\simeq c_s$, where $c_s$ is the propagating speed of adiabatic perturbation also known as the speed of sound, given by
\eq{
c_s=\left(1+\frac{4\Omega_1^2}{M^2}\right)^{-1/2}
}
where $M^2$ is given by,
\begin{align}
M^2= Q_2+\eps_H H^2 \mathbb{R}-\Omega_1^2,
\end{align}
with the quadratic form $Q_2$ given by 
\eq{
Q_2=&e_2^a e_2^b V_{ab},
}
where $e^a_2$ are the normal unitary directions in the Frenet-Serret frame, and $\mathbb{R}=\mathbb{R}_{abcd}e_1^a e_2^b e_1^c e_2^d$. Notice that, the maximum and minimum values of $Q_2$ are given by the eigenvalues of the Hessian matrix $V_{ab}$ denoted $\lambda_{\text{min}}$ or $\lambda_{\text{max}}$, with  $\mathbb{R}_{abcd}$ and $\mathbb{R}$  the Riemann tensor and Ricci scalar of the scalar field manifold. \\

Hence the Lyth bound  (\ref{use2}) can be written as, 
\eq{
\Delta \phi\geq \kappa\sqrt{\epsilon_H}\left(1+\frac{4\Omega_1^2}{M^2}\right)^{1/4}=\kappa\sqrt{\epsilon_H}\left(1+\frac{4\Omega_1^2}{\lambda-\Omega_1^2+\epsilon_HH^2\mathbb{R}}\right)^{1/4},
\label{use3}
}
where $\lambda\in [\lambda_{min},\lambda_{max}]$. \\

Let us also comment on different scenarios, by considering the CEB (\ref{gamma}),
\begin{enumerate}
\item
If $c_s\simeq 1$, it looks like the bound can be made even stronger than in (\ref{use1}). This is because $c_s=1$ implies that either $\Omega_1=0$ $-$which holds for the case of a single field$-$ or $M^2\gg \Omega_1^2$. For $\epsilon_H\ll 1$ the last case implies that $\lambda$ cannot be negative, indicating that the second inequality of the dSC  does not hold and that $\lambda\gg \Omega_1^2$.  Also
for inflation to take place, we  need  $\gamma\ll 1$ which makes the bound for $\Delta\phi$ stronger  compared to the bound fixed by $\epsilon_H$.
\item
If $c_s<1$, $\Omega_1$ cannot be zero. This case then applies to multi-field scenarios.  Notice that  the bound on $\Delta\phi$ is now weaker than the bound for single field models (\ref{use1}) if $\lambda<\Omega_1^2$. 
As shown in \cite{Scalisi:2018eaz},  if $\Delta\phi\ge 10$ in Planck units, there is still room to satisfy the dSC. So now we have to see under which conditions the ratio $\sqrt{\gamma/c_s}$  remains unaltered, 
allowing the dSC to be satisfied.  If $\lambda-\Omega_1^2>0$ the bound can become stronger.
\end{enumerate}

An extra interesting implication of the Lyth bound is the posibility to construct an upper bound on the field displacement $\Delta \phi$  under the assumptions of the previous section.  Since  $\epsilon_H\geq \gamma$, expression (\ref{use3}) can we rewritten for $\lambda_{min}$ as
\begin{equation}
\lambda_{min} \geq \frac{\kappa^4\gamma^2 (\epsilon_HH^2\mathbb{R}+3\Omega_1^2)-\Delta\phi^4(\epsilon_HH^2\mathbb{R}-\Omega_1^2)}{(\Delta\phi^4-\kappa^4\gamma^2)},
\label{lytha}
\end{equation}
where we have assumed $\Delta\phi>\kappa\sqrt{\gamma}$. By comparing innequalities (\ref{lytha}) and (\ref{lambda}), we obtain
\eq{
\Delta\phi\leq
\kappa\sqrt{\epsilon_H}\left(1-\frac{4\Omega_1^2}{\frac{3H}{2}D_t\log\gamma-3H^2(2\epsilon_H-\omega'+\frac{1}{3}\epsilon_H\mathbb{R})}\right)^{1/4}.
}
Let us consider the case in which $\gamma\leq\epsilon_H\ll 1$ with $\omega'\ll 1$.
By assuming a non-zero but very small $\gamma$ (such that $D_t\log \gamma$ is much larger than $\epsilon_H$) it follows that it is possible to have an upper bound for the field displacement $\Delta\phi$. The  bounds for $\Delta\phi$ now read 
\eq{
\kappa\sqrt{\epsilon_H}\left(1+\frac{4\Omega_1^2}{\lambda_{min}-\Omega_1^2}\right)^{1/4}\leq \Delta\phi\leq \kappa\sqrt{\epsilon_H}\left(1-\frac{8\Omega_1^2}{3HD_t\log\gamma}\right)^{1/4}.
}
This result is consistent only if $\lambda_{min}\leq -(3H/2)D_t\log\gamma +\Omega_1^2$ as implied by (\ref{lambda}). It then follows that there must be a bound for  field displacement,  in agreement with the Swampland Distance Conjecture. In particular we observe that trans-Planckian displacements for the field require large values of $D_t\log\gamma$.\\

In summary, by assuming a non-zero contribution to the entropy (from some fundamental quantum quantities connected to  string theory), we see that it is possible to obtain the dSC with explicit expressions for the constants $c$ and $c'$ depending on the entropy $\gamma$, the Hubble constant $H$, the curvature radius $\Omega_1$ and the usual slow-roll inflationary parameters. Importantly enough, we can also reproduce the mutually exclusive inequalities in the dSC.\\

We now proceed to consider a simple toy example in which the entropy contribution from a flux string scenario is computed.

\subsection{Entropy from a flux compactification on $T^6/{Z}_2$}
\label{}

Let us study a simple flux compactification on an isotropic six-dimensional torus threaded by fluxes and in the presence of an orientifold $O3$-plane. We are not considering the presence of $D3$-branes for which we are taking RR potential $C_4=0$.  We will follow the notation of \cite{Kachru:2002he} in this subsection.\\

\begin{figure}[!htb]
\centering
\begin{tikzpicture}
\draw[thick,black] (0,0) -- (0,2) -- (2,2) -- (2,0) -- (0,0) ;
\draw[thick,black] (4,0) -- (4,2) -- (6,2) -- (6,0) -- (4,0) ;
\draw[thick,black] (8,0) -- (8,2) -- (10,2) -- (10,0) -- (8,0) ;
\node at (1,-0.3) {$y^1$} ;
\node at (-.3,1) {$y^2$} ;
\node at (-.3,1) {$y^2$} ;
\node at (5.0,-0.3) {$y^3$};
\node at (3.7,1) {$y^4$} ;
\node at (3.0,1) {$\times$} ;
\node at (9.0,-0.3) {$y^5$};
\node at (7.7,1) {$y^6$} ;
\node at (7.0,1) {$\times$} ;
\end{tikzpicture}
\caption{Coordinates of the compact space.}
\label{fig:torus}
\end{figure}
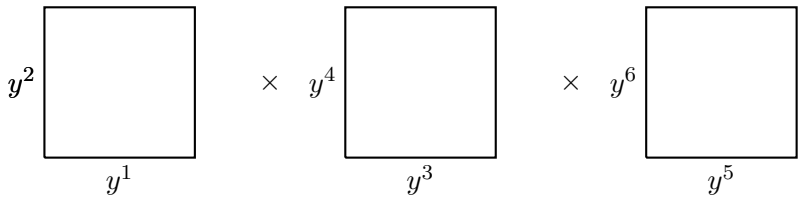

Consider a factorizable torus $T^6=T^2\times T^2\times T^2$ with coordinates as in Fig. \ref{fig:torus},\\
 
and let us take the ten-dimensional metric given by 
\eq{
ds_{10}^2 = ds_4^2 + \sum_{i} (dy^i)^2 \,,
}
where $ds_4^2$ is the homogeneous and isotropic FRW metric (\ref{FRW}). Since we are interested in computing an amplitude transition of a given flux configuration of $N$ units of RR flux $F_3$ and $M$ units of NS-NS flux $H_3$  which satisfy the supergravity equations of motion and  the stringy constraints, we shall consider that during transition the fluxes are time dependent.  Let us start by considering a specific example  in which the closed string potentials $C_2$ and $B_2$ depending on time, are given by
\begin{equation}
C_2=f(t)Ny^6dy^2\wedge dy^4, \quad \text{and} \quad B_2=f(t)My^5dy^1\wedge dy^3,
\end{equation}
because the function $f(t)$ depends on time. The corresponding action in  the string frame
\begin{equation}
\mathcal{A}_{\text{flux}}=-\frac{1}{2}\int d^{10}x\sqrt{-g_{10}}~\left(F_3^2+ e^{-2\phi}H_3^2\right),
\label{fluxaction}
\end{equation}
reduces to
\begin{equation}
\mathcal{A}=\frac{1}{2}\int d^{10}x ~a^3(t)\left[ N(\dot{f}(y^6)^2-f^2)+Me^{-2\phi}(\dot{f}(y^5)^2-f^2)\right]
\end{equation}
where the 3-form fluxes are defined as
\begin{eqnarray}
F_3&=&N(\dot{f}y^6 dt+fdy^6)\wedge dy^2\wedge dy^4,\nonumber\\
H_3&=&M(\dot{f}y^5dt+fdy^5)\wedge dy^1\wedge dy^3,
\end{eqnarray}
Since we are considering the isotropic case, we take all the internal coordinates $y_i$ on equal footing $(y)$, and the above integral can be written as\\

\begin{equation}
\mathcal{A}_{\text{flux}}=\frac{1}{2}\int d^{10}x~a^3(t)(N+e^{-2\phi}M)\left( \dot{f}(t)y^2-f(t)^2\right),\\
\end{equation}
which mimics the behavior of a harmonic oscillator in a  compact space. Thus the partition function related to the above action can be found explicitly. The first step is to  integrate over the internal coordinates from 0 to $2\pi R$ and external space coordinate $r$  from $0$ to $r_\ast$, for some finite $r_\ast$ from which one obtains
\begin{equation}
\mathcal{A}_{\text{flux}}=-\int_0^{t_\ast}dt~f(t)\hat{D} f(t),
\end{equation}
where
\begin{eqnarray}
\hat{D}&=&\alpha(t)\left(\frac{d^2}{dt^2}+3 H \frac{d}{dt} + \omega_0^2\right),
\end{eqnarray}
with $\omega_0^2=3/(2\pi R)^2$ and
\begin{equation}
\alpha(t)=\frac{(2\pi)^9 R^8 r_\ast^3}{9}a^3(t)(N+e^{-2\phi}M),
\end{equation}
with the boundary condition $f(0)=f(t_\ast)$, which states that at time $t=0$ there is a given distribution of fluxes with  $N$ units of RR flux and $M$ units of NS-NS flux. Between time 0 and $t_\ast$ this distribution of fluxes changes such that at time $t_\ast$, the original distribution of fluxes is recovered. It is important to notice that $\alpha(t)$ is finite due to the tadpole cancellation condition and the fact that  the dilaton has been fixed by the fluxes.\\

Therefore the partition function can be given by

\begin{equation}
\mathcal{Z}_{(N,M)}(0,t_\ast)=\int \mathcal{D}[f(t)]e^{-\frac{i}{2}\int_0^{t_\ast} dt~\alpha(t) f(t)\hat{D}f(t)},
\label{411}
\end{equation}
which can be used to compute the transition amplitude between an initial flux configuration $(N,M)$ and the final one specified by the same flux configuration. After performing the integral of (\ref{411}) (and replacing $t_\ast$ by $t$), and by taking the case in which $\Omega^2= \frac{9H^2}{4}-\omega_0^2>0$, one obtains 
\begin{equation}
\mathcal{Z}_{(N,M)}(0,t)=\left(\frac{-i\alpha(t)\Omega}{2\pi \text{sinh}(\Omega t)}\right)^{1/2}.
\label{Z}
\end{equation}
Notice that, our model behaves like a harmonic oscillator with a time-dependent mass $\alpha(t)$. \\

Now, after using standard methods in Euclidean time, it is possible to compute 
the entropy $S(t)$ from $\mathcal{Z}_{(N,M)}$  which reads
\begin{equation}
S(t)=\frac{1}{2}\log \left[-\frac{i}{2\pi}\alpha(t)\Omega \text{csch}(\Omega t)\right]+\frac{1}{2}\left(\dot\Omega t+\Omega \right)t ~\text{coth}(\Omega t)-\frac{t}{2}\left(\frac{\dot{\alpha}}{\alpha}
+\frac{\dot\Omega}{\Omega}\right)
\end{equation}
where
\begin{equation}
\frac{\dot\alpha}{\alpha}=3H, \qquad \text{and} \qquad \frac{\dot{\Omega}}{\Omega}= \frac{9}{4}\frac{H\dot{H}}{\Omega^2}=-\left(\frac{3H}{2\Omega}\right)^2 H\epsilon_H,
\end{equation}
while time variation of $S$ is given by
\begin{eqnarray}
\dot{S}&=&-\frac{1}{2}\left[ \Lambda H\epsilon_H t-1\right]^2\Omega^2 t ~\text{csch}^2(\Omega t)\nonumber\\
&&+\Lambda\epsilon_H H\left[\frac{H}{2}\left((\epsilon_H+2\eta_H)-\Lambda\epsilon_H\right)t-1\right]\Omega t ~\text{coth}(\Omega t)\nonumber\\
&&+\frac{1}{2}\left[3\epsilon_H+\Lambda\epsilon_H(2\Lambda \epsilon_H -\epsilon_H-2\eta_H)\right]H^2 t,
\end{eqnarray}
where $\Lambda=\left(\frac{3H}{2\Omega}\right)^2$.  In the following we  are going to consider appropriate limits in order to see if the obtained value for $\gamma$ is consistent or not, meaning a positive value for $\dot{S}$ for some time interval.

\subsubsection{Swampland implications}
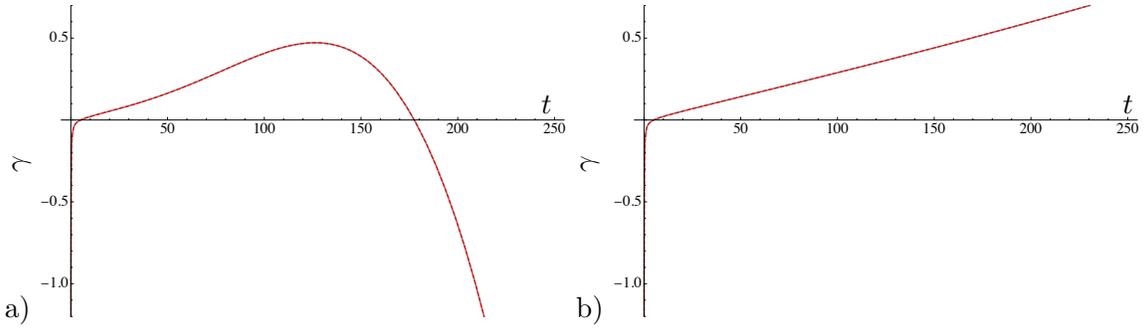
\begin{figure}[htbp]
   \centering
a) \includegraphics[width=7cm]{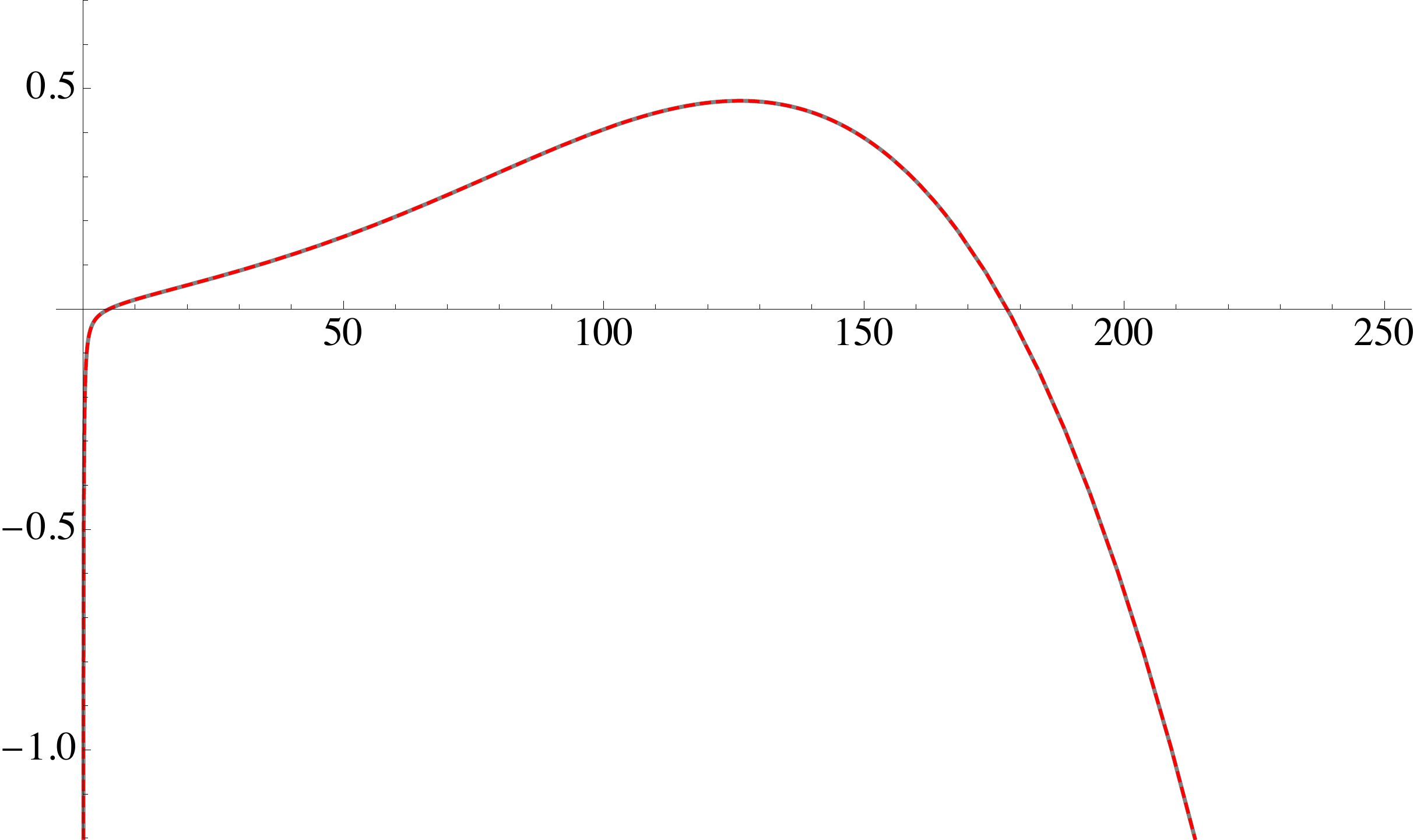} b) \includegraphics[width=7cm]{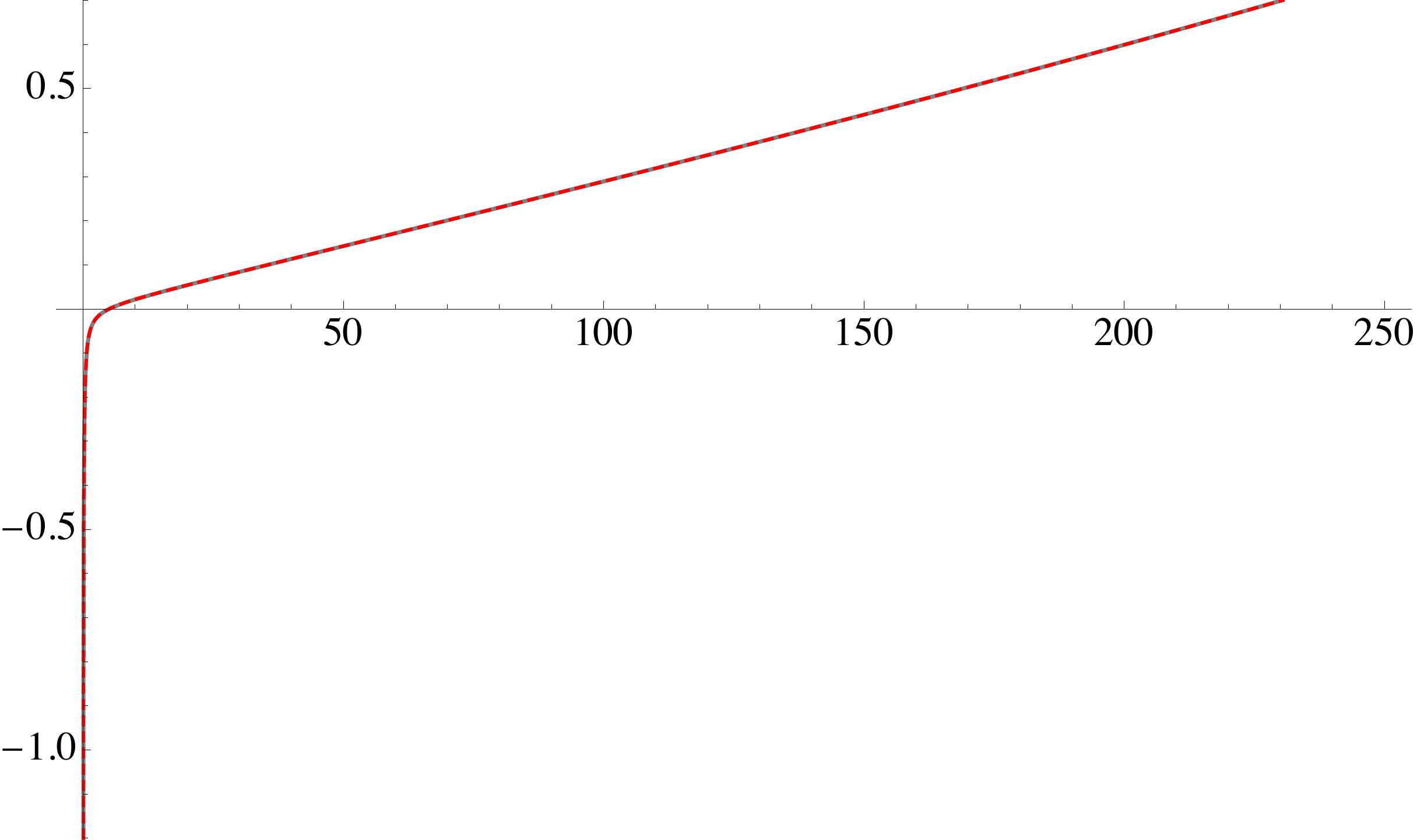}
\begin{picture}(0,0) 
   \put(-10,77){$t$}
   \put(-230,77){$t$}
   \put(-430,55){\rotatebox{90}{$\gamma$}}
    \put(-215,55){\rotatebox{90}{$\gamma$}}
  \end{picture}
\caption{ \label{fig:gamma} By taking a small Hubble constant, $H_{inf}<2.5 \times 10^{-5}$ in units of Planck mass \cite{Akrami:2018odb} and small values for the internal volume we plot $\gamma$ as a function of time for $\Omega\ll H$ or equivalently as in (\ref{Lambda}) for $\Lambda\gg 1$ implying small $\Omega$.  a) $\Omega^2=  \left( \frac{3 H}{2} \right)^2 -\omega_0^2> 0$ for which $\gamma$ depends on hyperbolic trigonometric functions  and b) $\Omega^2 <0$ for which $\gamma$ depends on trigonometric functions. In both cases $\Omega$ large leads to negative values for $\gamma$.}
\label{figura}
\end{figure}

Remember we are interested in studying inflation which requires the conditions,  $\epsilon_H, \eta_N \ll 1$ and check if the model is compatible with CEB by obtaining $\gamma\ll 1$. For the case we are considering, the only limit that makes sense is when one takes a very small value of $\Omega$ which implies that\footnote{Notice that $\Omega$ cannot be arbitrarily large.} $\Lambda$ is very large and that the volume of the internal manifold produces a lower bound given by
\begin{equation}
 \text{Vol}(T^6)\gtrsim \frac{1}{27 \pi^6 H^6},
\end{equation}
implying that the internal volume is approximately given by the Hubble radius during an accelerated expansion.  Now, we are taking a special limit in which
\begin{equation}
\Lambda\sim \frac{1}{\epsilon_H},
\label{Lambda}
\end{equation}
for which $\gamma$ is  given by
\begin{equation}
\gamma\approx -\frac{H}{4\pi}(Ht-1)^2\Omega^2 t~ \text{csch}^2(\Omega t)-\frac{H^2}{2\pi}(1+\frac{Ht}{2})\Omega t ~\text{coth}(\Omega t) + \frac{H^3}{2\pi} t.
\end{equation}
As depicted in Fig. \ref{fig:gamma}, $\gamma$ is positive and smaller than 1 for very large values of $\Lambda$, making our whole model consistent. This means, as we have witnessed in section 3, that we can reproduce the dSC and the mutual exclusiveness among the corresponding inequalities.\\

Since we are dealing with an effective single field inflationary  scenario,  from (\ref{1des}) the first inequality in the dSC reads
\begin{equation}
\left(\frac{V'}{V}\right)^2\geq -\frac{H}{2\pi}(Ht-1)^2\Omega^2 t~ \text{csch}^2(\Omega t)-\frac{H^2}{\pi}(1+\frac{Ht}{2})\Omega t ~\text{coth}(\Omega t) + \frac{H^3}{\pi} t,
\end{equation}
which for small $H$ can be written as
\begin{equation}
\left(\frac{V'}{V}\right)^2\geq a H\sim a V^{1/2},
\end{equation}
where $a$ is derived from the $\gamma$ expression.
Therefore, for this case, since $\gamma<1$, eternal inflation seems to be allowed and a quantum breaking time is actually happening for $t\sim 1/H^3$ \cite{Dvali:2018jhn}. Notice also that while the first inequality in the dSC is not fulfilled, the second one  is and in consequence inflation cannot last for sufficient amount of time.\\

For the second case in which $\Omega^2<0$  the volume of the internal manifold is bounded as
\begin{equation}
\text{Vol}( T^6)\lesssim\frac{1}{27\pi^6 H^6}.
\end{equation}
We get a similar expression for $\gamma$ in the limit of small $\Omega$ with the hyperbolic trigonometric functions replaced by trigonometric ones. In this case, there is also a time interval at which $\gamma$ is positive, but not restricted to be small (see Fig. \ref{figura} b). Therefore, there are zones with $\gamma$ of order 1 and for later times, $\gamma>1$.  While $\gamma<1$ inflation is allowed, but when we reach times at which $\gamma$ is of order 1 or higher, inflation stops. Then, we can see that entropy produced from fluxes is a natural mechanism to stop an accelerated expansion of the Universe.


\section{Conclusions and Final comments}
The de Sitter Conjectures (dSC) have stringent implications on four-dimensional effective  field theories derived from a theory of quantum gravity such as string theory. For instance, if the conjectures are true,  single field inflationary models are ruled out. However, the situation is comparatively  optimistic for the multi-field cases. Given the implications of these conjectures on the understanding of our Universe, it is desirable to relate them to other well-tested conjectures like the Covariant Entropy Bound, involving some  fundamental microscopic features of the Universe. This will test those conjectures from a more fundamental view point and put them in a stronger ground.\\

By using the Covariant Entropy Bound on an early phase of a four-dimensional accelerated universe, we present a bound on the slow-roll parameter $\epsilon_H$ in terms of the time derivative of the entropy,  given  by $\gamma< \epsilon_H$.  In a thermodynamically closed four-dimensional Universe on which $\gamma \approx 0$, there is no obstruction for inflation to occur. However,  by extending this effective model to include extra dimensions (indicating that we have drastically modified the theory to be compatible with string theory), it is possible to consider an extra entropy source. In this case, $\gamma$ does not necessarily vanish, opening up the possibility to either compromising inflation or establishing some constraints in the slow-roll parameters such as $\epsilon_V=\nabla V/2V$.\\

In fact, for the single-field models we find that $\epsilon_V<\sqrt{2\gamma}$,  ruling out inflation for $\gamma$ of order 1 or higher in concordance with the dSC. A simple toroidal flux compactification shows that fluxes constribute to the four-dimensional entropy while fulfilling the dSC.\\

For the multi-field scenario, we reproduce (in the Frenet-Serret frame) both inequalities in the de Sitter Conjecture (\ref{SdSC}) as well as the fact that they are mutually exclusive. For $\gamma \approx 0$,  we reproduce the result first presented by Ach\'ucarro and Palma \cite{Achucarro:2018vey}. For $\gamma\ne 0$, we provide an expression for $c$ and $c'$ in terms of physical quantities in the effective field theory such as $H$, $\gamma$ and the field dynamics, whose values determine the order of the constants.    Besides, in agreement with the distance conjecture, the length displacement of a scalar field is bounded from above, where the limiting bound is described in terms of entropy changes. In this sense we observe that  all these swampland criteria are compatible with the CEB.\\


\begin{center}
{\bf Acknowledgments}
\end{center}

\noindent
We thank Bruno Valeixo Bento for his observations and suggestions on our manuscript and to Nana Cabo, Yessenia Olguin and Ivonne Zavala for very useful discussions. The work of D.C  and A. G B was partially supported by a CONACyT graduate scholarship. C. D and  O. L-B are partially supported by CONACyT project  CB-2015-258982.

\appendix

\section{Frenet-Serret frame}
\label{A1}
In order to have a more general description we  shall consider $N$ fields. For that, let us define a map
\begin{equation}
\vec{\phi}:\mathbf{I}\rightarrow \mathbf{R}^n,
\end{equation}
such that
\begin{equation}
\vec{\phi}(t)=(\phi^1(t),\dots ,\phi^n(t))
\end{equation}
where the entries are given in terms of a time independent basis $\hat{u}_a$, i.e.,
\begin{equation}
\vec{\phi}(t)=\phi^a(t)\hat{u}_a,
\end{equation}
with 
\begin{equation}
||\vec\phi(t)||^2=\phi^a\phi^bh_{ab}=\phi^2,
\end{equation}
where $h_{ab}$ is the metric of the moduli space we are assuming  to be explicitly independent of time.\\

As time runs, the trajectory in the field space has a tangential vector defined as
\begin{equation}
\hat{e}_1(t)=\frac{\dot{\vec{\phi}}}{\dot\phi}=\frac{\dot\phi^a}{\dot\phi}\hat{u}_a=T^a\hat{u}_a,
\end{equation}
with
$\dot{\vec{\phi}}(t)=\dot{\phi}^a(t)\hat{u}_a$
and $||\dot{\vec{\phi}}(t)||^2=\dot\phi^a\dot\phi^bh_{ab}=\dot\phi^2$.  The normal unitary vector can be constructed as follows. Consider
\begin{equation}
\ddot{\vec{\phi}}(t)=\ddot\phi^a(t)\hat{u}_a,
\end{equation}
and the normal vector $\vec{e}_2(t)$  given by
\begin{equation}
\vec{e}_2(t)=\ddot{\vec{\phi}}(t)-\langle\ddot{\vec{\phi}}(t), \hat{e}_1\rangle\hat{e}_1(t).
\end{equation}
After some calculations, the normalized normal vector is given by
\begin{equation}
\hat{e}_2(t)= N^a\hat{u}_a,
\end{equation}
with
\begin{equation}
N^a=\frac{\dot{\phi}^2\ddot{\phi}^a-\dot{\phi}^a\dot{\phi}^b\ddot{\phi}_b}{\dot\phi(\dot{\phi}^2\ddot{\phi}^2-(\dot{\phi}^b\ddot{\phi}_b)^2)^{1/2}}.
\end{equation}

Now, in terms of an orthonormal basis at each point of the curve $\vec{\phi}(t)$, i.e. the FS frame, we have that
\begin{equation}
D_t \begin{bmatrix}
\hat{e}_1\\
\hat{e}_2\\
\hat{e}_3\\
\vdots\\
\hat{e}_{n-1}\\
\hat{e}_n
\end{bmatrix}
=\begin{bmatrix}
0 &-\Omega_1&0&&&\\
\Omega_1&0&-\Omega_2&&&\\
0&\Omega_2&0&-\Omega_3&&&\\
&&&\ddots&&\\
&&&\Omega_{n-2}&0&-\Omega_{n-1}\\
&&&0&\Omega_{n-1}&0
\end{bmatrix}
\begin{bmatrix}
\hat{e}_1\\
\hat{e}_2\\
\hat{e}_3\\
\vdots\\
\hat{e}_{n-1}\\
\hat{e}_n
\end{bmatrix}
\end{equation}
and in consequence $D_t\hat{e}_1=-\Omega_1\hat{e}_2$. From the corresponding equations of motion 
\begin{equation}
D_t\dot{\phi}^a+3H\dot{\phi}^a+V^a=0,
\end{equation}
we can deduce the expression for $\Omega_1$,  where
\begin{eqnarray}
D_t\dot{\phi}^a&=&\ddot{\phi}^a+\Gamma^a_{bc}\dot{\phi}^b\dot{\phi}^c,\\\
V^a&=&\vec{\nabla}V\cdot \hat{u}_a.
\end{eqnarray}
and  $\hat{u}_a$ is a  time independent basis for the $N$-dimensional field space. \\

The $N$ equations can be written  in a vector notation as
\begin{equation}
(\ddot{\phi}+3H\dot{\phi})\hat{e}_1+\dot\phi (D_t \hat{e}_1)+\vec{\nabla}V=0,
\end{equation}
where $\hat{e}_1$ is the unitary tangent vector to the field trajectory. Since $D_t\hat{e}_1=-\Omega_1\hat{e}_2$, where $\hat{e}_2$ is the unitary normal vector to the trajectory, we have
\begin{equation}
(\ddot\phi+3H\dot\phi)\hat{e}_1-\dot\phi\Omega_1\hat{e}_2+\vec{\nabla}V=0,
\end{equation}
Now we want to express components of $\vec{\nabla}V$ in terms of the FS basis, this is
\begin{equation}
\vec{\nabla}V\cdot\hat{e}_j(t)=V_ae^a_j=V_j,
\end{equation}
where we have used $a$ as an index for the basis $\hat{u}_a$ and $j$ for the index in the FS system.

\bibliographystyle{JHEP}
\bibliography{References}  

\end{document}